\documentstyle [12pt]{article}
\evensidemargin\oddsidemargin
\begin{document}
\count0 = 1
\vspace{30mm}
 \title{{ Fuzzy Phase Space  Structure \\    
  as Approach to Quantization \\ }}
\author{S.N.Mayburov \\
Lebedev Inst. of Physics\\
Leninsky Prospect 53\\
Moscow, Russia, 117924\\
Email: mayburov@sci.lpi.msk.su
\\}
\date { }
\maketitle

\begin{abstract}

Modification of  a nonrelativistic phase space
  structure based on fuzzy ordered sets (Fosets) structure 
investigated as a possible nonrelativistic quantization framework.
 In this model
 particle's $m$ state corresponds to Foset element - fuzzy point.
Due to fuzzy ordering its space coordinate 
$x$ acquires principal uncertainty $\sigma_x$.
It's shown that  proposed Mechanics on fuzzy phase space manifold 
reproduces the main quantum effects, in particular
the  interference of quantum states.  

\end{abstract}
\small{Talk given on 'Quantum Foundations' conference\\
Vaxjoo, Sweden, june 2001, to appear in Proceedings\\}
\section { Introduction}

Quantum Mechanics (QM) is now a well established physical theory,
yet its relation with physical space-time structure and relativity isn't
 quite clear and actively discussed now. This interest enforced 
by the recent  indications that space-time properties at small (Planck) scale
can be quite exotic \cite{Aha,Dop}. Due to the absence of
 any experimental 
information it seems instructive to look for some indications 
reconsidering under this angle  standard Quantum Physics space-time picture.

The additional attention to QM foundations induced also 
 by  Gravity quantization problems. 
In particular Isham proposed that
 some  space-time properties like the metrics or topological structure,
assumed in the standard quantization should be modified significantly
or even rejected completely   \cite {Ish}.
Our work motivated largely by this ideas
in which framework we explore  Set structure of space-time manifold $M_{s-t}$.
 Remind that for Euclidean Geometry in 1-dimension
 its basic elements are points
$x_i$ which are ordered, as follows from Euclidean metrics definitions ;
 i.e.  proposition :
$$
 LP_x  : \quad \quad x_1 \le x_2.or.x_2 \le x_1
$$
 is true for arbitrary $x_1,x_2$
and thus $x_i$ set $X$ is the ordered set. Classical particle $m$ state
corresponds to $M_{s-t}$ point $x(t)$ and formally this state is the
point $r_p(t)$ in phase space.
In QM one regards as particle's  state the extended object - Dirac
 state vector $\Psi$ evolving on the same  $M_{s-t}$ manifold 
,i.e  classical space-time transferred to QM copiously.

Yet Set Theory permit other set structures, for which elements 
are weakly or fuzzy ordered relative to each other.
In our approach to quantization  such fuzzy ordered set (Foset)
constitutes the basic space-time manifold $M_F$
  with fuzzy relations between its  elements
 - fuzzy points for which proposition $L_x$ can be untrue
and only a weak propositions of the kind :'$x_1$ is in $x_2$ vicinity
of the approximate width $\sigma_x$'
characterizes a points relations. On this manifold 
  the particles evolves which are 'pointlike'  in a sense that their
 states corresponds to this  fuzzy points.
If the space coordinate axe $X$ related to some point $O$
 can be defined on $M_F$ , then  fuzzy points will be smeared
on $x$ axe with arbitrary dispersion $\sigma_x$. Furthermore
the particles mechanics  on fuzzy manifold described here
permit to reproduce
the  main quantum effects.

Some years ago it was shown that fuzzy sets formalism can
have important QM applications, in particular a fuzzy observables 
are the natural generalizations of QM observables \cite {Ali}.
Such formalism of 'fuzzy lumps' also was applied in quantum gravity
and cosmology studies \cite {Ren}. 
 In the last years it was shown
that some of fuzzy sets formalisms
 are appropriate also for Quantum Logics 
 (\cite {Pyk} and ref. therein). Yet it exploits
 nonstandard logics
of propositions with multiple outcomes , besides standard yes/no.

  Quantum  phase space geometry
to our knowledge never been investigated in fuzzy sets framework
and we try here to make first steps in this direction.
Recently the progress of noncommutative geometry and its 
 applications \cite {B}
enforced interest to  space-time geometric structure
\cite {Con,Dop}. It supposes that space-time coordinates $x^i$
are noncommutative which can be revealed at Planck scale,
which as will be shown intersects with fuzzy theory results.
Note that the term 'Fuzzy Geometry' sometimes used for
Nocommutative Geometry ideas, but here we'll use it in its proper 
meaning for Foset based manifold.  
This Fuzzy Geometry  approach to QM can be interpreted
 as the novel quantization
formalism i.e.  formal transition from Classical mechanics to QM which
was formulated already via path integrals, Von Neuman algebra,
deformations and other approaches.
In our case the quantization regarded as transfer from Classical
ordered phase space to fuzzy one.

We don't present here the complete theory of systems evolutions on
fuzzy manifolds, rather this text is semiqualitative discussion 
of the main ideas.
Any  quantization formalism inevitably intersects with QM
interpretation problems, discussed here in the final chapter.
For our study the most interesting seems the physical meaning
of a quantum state vector and in particular the role of complex
numbers in QM.

Any novel QM formalism must include the description of its tests
in corresponding measurements.
The transition from quantum states to experimental
probabilities $P(x)$ is the subject of quantum  measurement theory,
which up to now doesn't have acknowledged formulation \cite {Busch}.
 Thus one should use some variant of 
quantum measurement theory to make  formalism fully consistent.
But as will shown below our model is  indifferent to
their choice, but most preferable are models which literally
use Schrodinger dynamics like MWI and decoherence models \cite {Busch}.
Of them the most preferable seems dual selfdescription
formalism \cite {Mayb8}. 
In particular it's important to understand in principle how
transition from $w^Q$ to random events (state collapse) occurs.
In our paper measurement aspects aren't regarded,
restricting only to statistical quantum ensembles for which
it's enough to use $W^Q$. We also use some measurement theory
methods  alike comparison of mixed and pure states distributions.

\section {Fuzzy sets and relations}

Let's remind the main properties of fuzzy sets and fuzzy relations.
Beside standard 'crisp' sets in Abstract Algebra and Set Theory
 fuzzy sets $A^F$ also regarded for which verity of
propositions alike $a_i \in A^F$ characterized by positive weight 
$w_i \ge 0$, in place of $1,0$ in standard logics \cite {Zad}.
 Here $a_i$ are elements of standard set $A$.
  We'll regard in our model fuzzy
relations $R$ and mappings on standard sets,
and thus don't need to describe fuzzy sets properties in
details, which can be found elsewhere \cite {Got}.


Note that any geometry based on topology and set theory premises which
stipulates its manifold properties. 
For example  1- dimensional Euclidean  geometry set of elements
(points) is the   ordered set $X$ of continuum power.
 Thus relation of the
kind $x_i \leq x_j$ (or vice versa) always can be introduced
 for any  pair of its elements in 1-dimensional case,
for 3 dimensions they are analogous. Its topology is
compatible with differentiable manifold. If metrics also defined then
the distance $r_{ij}$ between  $x_i,x_j$ can be introduced.

To introduce
Fosets and the fuzzy relations $R_f$
 remind that in partial ordered set (Poset)
$P_A$ some of its elements (but not all in general) ordered by the relation
$a_i\leq a_j$, which obeys to standard rules \cite {Got}.
For example the  element $a_k\in P_A$ ordered relative to some $a_j$
but isn't ordered relative to some others
 $a_i$ - i.e. unrelated  denoted as $a_k \wr a_i$.
 Consider discrete Poset $P_A$ which  includes  'test' subset $A^0$
and 'reference' subset $A^1$. In $A^1$ all elements are ordered
 and  elements indexes grows
correspondingly to it so that  $a_i \le a_{i+1}$.
$A^0$ element $a'_0$  unrelated to some $a_i \in A^1$
and let's suppose that all this $a_i$ belongs to interval $[a_l,a_{l+m}]$,
 i.e.   $a_l\leq a_i, a_i\leq a_{l+m}$. $a'_0$ ordered to other $a_j \in A^1$
 and   $a_l \le a'_0, a'_0 \le a_{l+m}$, so that 
 $a'_0$ also belongs to interval $a_l,a_{l+m}$,
 but is 'smeared' inside it
 corresponding to weak $P_A$ structure.
  Described situation can be interpreted as 
 $a_0,a_i$ are approximately equal up to some arbitrary uncertainty 
$\sigma_a$. To introduce the quantitative measure
 for it one put in correspondence
to each $a_i$ the weight $w^0_i \ge 0$ with norm $\sum w^0_i=1$. Poset 
 $P_A$ on which the fuzzy measure $w^l_i$ defined is Foset \cite {Zad}.
As the result $w^0_i$ gives more detailed description  
of fuzzy relations between $P_A$ elements; for example if 
$w^0_j \approx 1$ and all others $w_i^0\approx 0$
 it means that $a'_0$ nearly coincides with $a_j$.
Standard  ordering corresponds to $w^i_j=\delta_{ij}$ for equal 
$a_i,a_j$ and $w^i_{i+1}=w^i_{i-1}=\frac{1}{2}$ for $a_i \in A^1$.
If $A^0$ consists of more than one elements analogous fuzzy
relations $w^j_i$ relative to $A^1$ elements can be defined for all of them.
A relations between $A^0$ elements should be defined additionally; they can be
  ordered or unrelated independently of their relations
to $A^1$ elements. Thus the Fosets  structure is more strictly ordered
then Poset one but weaker then for an ordered set. Thus the foset 
ordering structure is more strict than Poset one but weaker then
 the standard ordering.

Reference subset $A^1$ can be substituted
by continuous metricized subset (axe) $X$, and if $A^0$ properties conserved   
than for $a'_0 \in [x_l,x_m]$ interval (which can be also infinite)
 the fuzzy $a'_0,x_i$ relations
described by continuous distribution $w^0(x)\ge 0$ with norm $\int w^0 dx=1$.

If fuzzy point $a'_0$ confined inside  interval $E_x=[x_a,x_b]$
 where $w(x)>0$
then $a'_0$ fuzziness expressed by the proposition
 $L_0$ which will be important for Fuzzy  Mechanics :
$$
 L_0 : \quad  \forall \Delta x_i \in E_x ; \quad a'_0 \in E_x \cap a'_0
 \notin \Delta x_i
$$ 
where $\Delta x_i \in E_x$ is any interval contained inside $E_x$.
 Regarded example is most simple and more complicated
fuzzy geometric relations exists; for our topics
most interesting  is the situation when $a'_0\in E_x$ which
consists of the 
two or more noncrossing intervals $Dx_n=[x^n_l,x^n_m]$ and in this
case $L_0$ is true also.
Generalization of this $A^0,X^1$ fuzzy relation from 1 to$n$-dimensions
is straightforward and doesn't contain any principally
new features. 
It's possible also to regard continuous fuzzy subset $X^0$  replacing
discrete $A^0$, but this case is physically uninteresting.
The generalization of described formalism to $n$-dimensions
is straightforward and don't include a new features important for us
\cite {Got}.
Formal definition of fuzzy relations  is the generalization
of regarded   examples and can be found elsewhere \cite {Got}.

\section {Fuzzy Mechanics (FM) and Fuzzy States}

Usually system quantum state described by
Dirac state vector presented as a complex function $\Psi(x)$ or
$\varphi(q)$ in arbitrary $Q$ representation.
 But operationally relevant
are a nonnegative distributions $w(x)=|\Psi(x)|^2$ or
correspondingly $w^Q(q)$ for other observables. Only them or $Q,Q'$
correlations are observed in the  experiments and from them
a system state vector $\Psi$ derived. 
In the real experiment they are realized via statistics of outcome of 
individual events occurs with probabilities $P(x)=w(x)$, etc..
Note that a quantum state $\Psi$ can be formally expressed via this finite
or infinite $\{w^Q\}$ set  $W^Q$ covering all possible system observables $Q$ 
 distributions (\cite {Busch} and ref. therein). This set regarded as the
special 'empirical'  
 representation of a quantum states which  regarded in our study
due to its principal importance 
despite its practical inconvenience.
 In some cases a state vector can be restored from
the restricted subset of this set.
 For example a spin$\frac{1}{2}$ state $\psi(s_z)$ can be restored
from $w(s_z),w(s_y),w(s_x)$ despite that the complete set $W^Q$ is infinite.
  

Now we discuss the transition from Fuzzy geometry to Fuzzy mechanics (FM),
analogously to transition from Euclidean Geometry to Classical Mechanics.
In such classical case the instant  position of the  particle is the point in
 Euclidean 3-space;
 its physical state corresponds to the point in the phase or
configuration Euclidean 6-space.
 Our main studied system is a massive nonrelativistic
 particle moving freely or in some potential field.
In our model  we identify a particle $m$ with the fuzzy point 
$a'_0$  regarded in the previous chapter.
As follows from described Foset properties, in fuzzy phase space
one can define coordinate axes $X,P$ associated with some material object $O$
chosen as reference frame (RF) and so $m$ is fuzzy point relative
to this ordered axes.
 Without an evolution in static situation $m$ 'state' 
supposedly defined by $w(x)$ relative to ordered axe $X$ completely.
If the evolution turned on then  such object - the fuzzy point $m$
 evolves between
the configurations where $x$ value is uncertain inside some interval
$[x_a(t),x_b(t)]$ as described by $w(x,t)$.
 Thus  $m$ velocity $v_{x}$ in general can't have 
 certain value and  supposedly 
 described by analogous fuzzy parameter with
the  distribution $w^v(v_x,t)$ on  $V_x$ velocity ordered axe associated with
$O$ RF. Particle $m$ characterized by its fixed mass parameter $m$
and in this fuzzy $X$ space can move freely
or in potential field $U(x)$. For the simplicity at this stage
 we suppose without proof that a  fuzzy  momentum $p$
is proportional to the given fuzzy $v_x$ 
 in a sense that $w^v(v_x)=w'(mv_x)$, where $w'$ is $p$ distribution.   
The principal uncertainty $\sigma_x$ of $m$  space coordinate $x$,
 velocity $\sigma_v$ and  momentum $\sigma_p$
indicates that FM should have some resemblance with QM and  we'll show
that $m$ evolution in this theories also has other common features.
 It supposed below that
$w(x),...,w^Q(q)$ distributions by means of some
experimental procedure can be measured. 

We'll suppose that  $m$ physical instant state in an arbitrary RF
  described by a 'fuzzy' state $ |g\}$ which account all $m$
 observables description. For the start
we make the  minimal assumptions about $|g\}$  as mathematical object,
to permit the  maximum range of possibilities
 and  try to derive its properties
from a fuzzy relations in the phase space.
$|g\}$ set denoted $M_s$; it doesn't supposed to be
 a normalized linear space  $M_L$ $a\quad priory$;
our aim is to study whether $M_s$ has such properties.
 Naturally  $|g\}$  have positive constant norm :
$$
   N=\|w\|=\int wdx=1
$$
$w(x)= F_x(g)$ is a real positive
 function on $X$ and some unknown functional of $|g\}$.
 Alike a state in any theory $|g\}$ should contain a complete information
 about  an arbitrary $m$ observable $Q$  distribution $w^Q(q)$ and
 if different $Q_i,Q_j$
related by some constraints or correlations it also should be accounted by $g$.
Remind that in QM $x,p$ distributions are correlated via the commutation
 relations.
$w(x)$ alone can't describe  a future $g$ evolution which can depend on 
$w'(p)$, etc.  and it must include the additional
  components denoted as $\bar{g}^x$, so that symbolically
$g=w \otimes \bar{g}^x$. 

Any $|g\}$ supposedly
can be decomposed into some arbitrary substates i.e. 'states parts',
 reflecting its fuzzy structure -  i.e a simultaneous
alternative possibilities coexistence in the phase space.
 For example consider  a fuzzy point  $m$ which belongs to
 the space region $E_x$ which consist of the
 two  noncrossing gaps $Dx_{1,2}$. If $m\in Dx_{1}$ only,
so that its weight in $Dx_2$ $w_2=0$  then 
$|g\}=|g'_1\}$. But if  both $w_{1,2} \ne0$
and $w_1+w_2=1$ then $|g\}=|g'_1\}\oplus|g'_2\}$
to which corresponds the following logical proposition for $m$ state :
$$
 L_E : \quad  m \in E_x \cap m \notin D x_1 \cap m \notin D x_2
$$
We don't define the summation rule $ad \quad hoc $ and can't calculate $g$ now,
meaning only $|g \}=F_s(g'_1,g'_2)$ i.e. $g$ is a general (nonlinear)
superposition which stipulates use of $\oplus$ sign. 
 Following the above arguments
each $g'_i$  should correspond to at least one 
'related' physical state $g_i$ with  the norm $1$    
which describes $m$ confined inside $Dx_i$ with the distribution
$w_i(x)=N_i^{-1} w(x)$.
 We'll regard more formal 
substates definition and their properties
in Appendix, but here it's enough to use
this semiqualitative description.
An analogous substates superpositions can be defined like states
in  the noncrossing intervals $[q_i,q_j]$ for other $Q$ observables.
In QM a substates consideration related closely with $\Psi$
 definition as a vector
in the linear Hilbert space but in FM it results from the
   $m$ properties as a fuzzy point and has less strictly defined 
features.

Obviously  that for regarded $L_E$ example the further $g$ decomposition
is possible - i.e.   $g'_i=\sum g'_{ij}$
for  any $Dx_{ij} \in Dx_i$
in which $Dx_i$ can be decomposed.
 Clearly such decomposition can be proceeded
to $g_{ijk}$, etc.. 
 In this approach one can consider the limit $Dx_i\rightarrow 0$
and represent 
 $g= \{ w(x); K(x,x',q,q'\}$   where $K$ correlation
tensor between different $x$ points and other phase coordinates $q$.
As the  example of the minimal FM theory
 which suits to Fuzzy Geometry picture we regard
that $m$ fuzzy state $g_F=\{w(x),K(x,x') \}$ where $K$ correlation
tensor doesn't depends of any $q$ and thus $\bar{g}^x=K$
  which together with $w(x)$
 defines a  future $g_F(t)$ state. Later we'll present more arguments
in favor of such $g$ structure, but here note only that in general 
such $g_F$ doesn't admit $x$-representation and only in the special
case discussed below it becomes possible.

We don't formulate yet evolution law for fuzzy states, but
at this stage  we'll use  simple  model with
  the simple assumptions formulated below. The first of them
is $w(x)$ norm conservation and corresponding $w$ flow equation
 should exist.
  Another one
is FM classical limit existence ; when $m$ mass is very large it
 becomes ordered localized point in phase space and
its trajectory defined by Hamilton equations with 
$H=\frac{P^2}{2m}+U(x)$.  
 Separately  should be considered assumption that $|g\}$ 
has $x$-representation $g(x)$ and $w(x)=F(g(x))$ and thus is local field
 but we don't assume it  at this stage.

$|g\}$ describes some extended object evolving in time and to reveal
its properties it's instructive to compare it with other extended
objects studied in Physics. Of them we'll consider here a classical
particle stochastic motion and a classical waves motion.
 In standard 
 classical statistical mechanics (CSM)
particle $m$ initial state is the random point in $R^6$ and described 
by probabilistic distribution $P (x,\dot{x})$.
Its evolution obeys to Classical Mechanics for an individual trajectory, but
 their ensemble characterized  by  probabilistic distribution
$P_e(x,\dot{x},t)$.
 As CSM state one can use $P(x,\dot{x})$ and $w(x)=P(x)$ is obtained by its
tracing. In fact  only  the quite simple CSM variant regarded here
and the real classical statistical
 theory permits much more complicated options. 
Until now all assumptions about $|g\}$ properties were applicable
both for CSM and FM, but now we come to their differences.
 CSM  evolution conserves the state norm and is additive for state
components.
 Additivity (linearity) means that if an initial state
 is the sum
 of two states $P_1,P_2$ with (probabilistic) positive  weights $r^p_1,r^p_2$ 
 then each  component evolves independently of other component presence 
the final distribution is:
$$
   P(x,t)=r^c_1 P_1(x,t)+r^c_2 P_2(x,t)
$$                        
In distinction as will be shown the pure fuzzy state $|g\}$ evolution
 due to the state structure (source) smearing (SS)
 effect can violate additivity and is nonlinear for $w(x)$.
 It constitutes the principal
 feature of fuzzy evolution and its  origin will be discussed here.
To illustrate it  consider the evolution of $m$ initial state 
$|g^0\}$  at $t_0$ on $X$  axe (1-dimension)
 which is the sum of two substates $g_{1,2}$  with $w^0_{1,2}(x)$
 disposed in the noncrossing  intervals
$Dx_{1,2}$. $m$ Initial state $g^0=g'_1\oplus g'_2$
 can be regarded as the source $S(g)$ for
the produced future state $|g(t)\}$ - signal, so that $w_s(x,t)=F_s(g^0,t)$.
 For the state $g(t)$ the following proposition analogous to
$L_E$ describes the fuzzy source structure at $t_0$:
$$
  L_f : \quad  S \in (Dx_1\cup Dx_2) \cap S
 \notin D x_1 \cap S \notin D x_2
$$
Suppose that $m$ state evolves freely and so that at some  $t$
distributions $w_{1,2}(x,t)$ for the independent  $g_{1,2}$ evolution
intersects largely. 
 What can be expected for the form of joint
distribution $w_s(x,t)$ ?
 For CSM states
such distribution will be additive sum $w^m_s=w_1+w_2$,
 because in any
individual event $m$ emitted by $Dx_1$ or $Dx_2$ separately and 
each $w_i$ corresponds strictly to one of this outcomes, i.e.
 is random source $S^m_R$.
But as follows from $L_f$ in case of fuzzy  $m$ source $S$ it
  can't be attributed to any of  $Dx_i$ separately. 
Due to it $w_s$ form should be such that it's in principle impossible
to  decompose it into the sum of two components
corresponding to $Dx_i$ sources. It should be maximally
different  from the mixture, so the $w_s^m$ content in $w_s$
 should be minimal. Due to it $w_s$
 can include principally nonclassical terms alike $\sqrt{w_1w_2}$ 
corresponding to $m$ source fuzzy position.  
Thus SS is the novel nonclassical feature of evolution 
and will be the main point of our attention throughout this paper.

 To study Fuzzy theory  we'll use for the comparison
  also probabilistic mixture $|g^m\}$ of 
several pure fuzzy states $|g_i\}$ and for them the additivity is fulfilled
 in some cases :
$$
   w^m(x)=\sum N_i w^r_i(x)=\sum w_i(x) 
$$
where, $N_i=\|w_i\|$; $w^r_i,w_i$ are normalized
 and unnormalized distributions correspondingly.

 We'll start FM study with  the simple
toy-model of fuzzy evolution and
as the  example  regard  
the two slits experiment (TSE) often used for a main QM effects discussion. 
It includes the pointlike source $E^m$ which emits in a wide cone
 a particles $m$ in the direction
of the flat screen
 with two parallel slits $\Delta x_{1,2}$ divided by the gap $2l_x$
( this set-up regarded to be 2-dimensional).
For simplicity $E^m$ supposedly emits the constant $m$ flow and 
 $w(x)$ doesn't depend on $t$. Due to it our problem
becomes analogous to the study 
  of a  fuzzy point $a'_0$   mapping on the 1-dimensional surface $X$. 
 Behind  this screen  at  distance $l_y$
the  photoplate $PP_x$ installed which measures $m$ coordinate $x$,
 (normally $l_x \ll l_y$). $\Delta x_i$  are very small;
 $\Delta x_i \ll l_x$ and  this slits 
can be regarded as pointlike sources $S_i$ for $m$ state on photoplate. 
It supposed  that the source $E^m$  produces on  the screen
 fuzzy state $|g^{in}\}$, which  passing through the slits  transfers into 
state $|g^0\}$   presented - i.e. $w^0(x)>0$ only in two  separated regions 
  $\Delta x_1,\Delta x_2$ centered around $x^0_{1,2}=\pm l_x$. Thus 
 $w^0(x)=w^0_1(x)+w^0_2(x); \quad w^0_1 \cap w^0_2=0$,
 corresponding to a fuzzy substates sum : $|g^0\}=|g^0_1\} \oplus |g^0_2\}$.

In our toy-model for TSE case   we'll assume that
all fuzzy effects depends only on  one fuzzy parameter  $I_s=\{1,2\}$
 - a number of slit. All other
$m$ evolution features supposedly are analogous to CSM.  
%
%
 Thus if only one slit $\Delta x_i$ is open $m$ final distribution
 will be the same as in particular CSM model. 
For the simplicity we choose $m$ spread to be spherically symmetric
relative to $m$ source - slit $S_i$   with $w_i(x,y)=w_e(\theta)w_r(r)$
 where $w_e$ is constant distribution.
It results in  $x$  distribution on the photoplate :
\begin {equation}
   w_i(x)=\frac{\|w_i^0\|}{(x^0_i-x)^2+l_y^2} \label {CC}
\end {equation}
 where the source intensity  is $N_j=\|w^0_j\|$.
Our final qualitative results for the fuzzy smearing effect
doesn't depend on exact $w_i$ form, but this ansatz is illustrative
because it's the monotonous function without zeroes.

 Denote $X_R=\Delta x_1 \cup \Delta x_2$.
If $m$ signal is a  probabilistic mixture $|g^m\}$ of two slits signals
its  structure described by the proposition for classical random source:
$$
   L_m :\quad   S^m_R \in \Delta x_1 \cup S^m_R \in \Delta x_2
$$
 from  which follows the distribution : 
$$ 
   w^m(x)= w_1(x)+ w_2(x)  
$$
is the additive sum of  the signals
 from the two slits.
%
 For a pure  initial fuzzy state from two slits 
 the proposition $L_p$ 
describes fuzzy source $S$ for TSE analogously  to $L_f$:
$$
   L_p :\quad   S \in X_R \cap S \notin \Delta x_1
 \cap S \notin \Delta x_2
$$    
The  distribution  $w_s(x)$ form depends on 
the source properties characterized by $L_p$ from which follows
 that for a fuzzy
 source    some $w_s$  positive component $w_n$
can't be attributed to $\Delta x_{1,2}$ individually
which responds to the regarded SS.  
Such problems studied in Information Theory and in particular
in Images Recognitions topics where fuzzy sets often applied
\cite {Kol,Got}.
The first problem is to define SS measure i.e.
 the separation criteria (SC) for the
discrimination of pure fuzzy states and mixture which 
can be ambiguous.
Note that from TSE description SC  applied to the situation when
the $m$ sources are very small $Dx_i\rightarrow 0$.
 Also we must find  the conditions under which
for a pure fuzzy state the complete SS can be achieved. For this purpose
let's define the measure of $w_1,w_2$ overlap $w_1 \cap w_2$ : 
$$
            C_x=\int \sqrt {w_1 w_2} dx
$$
In principle an alternative measures can be used, but for our problem
they lead to effectively same results. If $C_x \rightarrow 0$
TSE mixed and fuzzy distributions simply coincide.
 Our main observation  studied below is that
the achievement of maximal  SS for $w_s$ turns out to be the
severe constraint which in general permit to define  the principal
 $m$ evolution properties.

 The input-output flow conservation results in  $w^m,w_s$ norm equality :
$$
\|w_s\|=\|w^m\|=N_1+N_2=\|w_1\|+\|w_2\|
$$
, but $w_s$ form should  maximally
 differ from $w^m$, i.e. $w^m$ content in $w_s$ should be minimal as
$L_m,L_p$ propositions indicate. 
  $w_s$ can't depend only
on  each $w_i$ separately and must include some their nonlinear combinations
to become unrelated to any of this slit sources.  
Formulae for $w_s$ should be applicable also for one open slit
resulting in $w_s=w_i$ in this case, thus $w_s$ can be rewritten in the form :
\begin {eqnarray}
     w_s(x)=w_1(x)+w_2(x)+w_I(x) \quad  \\ \nonumber
  w_I(x)=F_I(w_1^{c_i}* w_2^{d_i})* F_c(V_g, x)
                                                            \label {CC1}
\end {eqnarray}
with $c_i,d_i>0; \, i=1,n$.
 $F_I$   can be decomposed as:
$$
      F_I=\sum a_i w_1^{c_i}* w_2^{d_i}  
$$
which can be a finite or infinite sum.
If we choose $F_c$ to be dimensionless, and no dimensional 
parameters contained in $F_I$, then  $c_i+d_i=1$. 
  $V_g$ denotes all other
$|g\}$ degrees of freedom (DF) except $w_i$. We suppose that the interference
 term (IT) $w_I=F_I*F_c$ admits such factorization,
 because $F_c$
which describes $m$ normalized distribution on $X$ naturally
  to be independent of
 the signal intensities (but not of other $|g\}$ parameters $V_g$).

 From  $\|w_s\|=\|w^m\|$  it results in $I_g=\int w_I dx=0$.
 $F_I$ is symmetric relative to $w_1,w_2$ and the simplest 
example is $F'_I \sim 2(w_1w_2)^{\frac{1}{2}}$.
 Of course other more complicated
$F_I$ should be regarded, but some illustrative calculations
will be performed for $F'_I$.
For the comparison note that in CSM $w_I=0$.

For illustration we start with the most simple
 and coarse example of SC.
$w_I$ can be negative at some $x$ and to consider only nonnegative
functions $w_n(x)$  which are nonadditive on $w_1,w_2$
one should add to $w_I$ some part of  $w_1+w_2$ defined by the condition
  $min(w_n)\ge 0$.
If to  present the  signal on the screen as :
$$
  w_s(x)=k_0 (w_1+ w_2)+[(1-k_0)(w_1+w_2) +w_I]=k_0 w^m+w_n
$$
with arbitrary $k_0\ge 0$ (it can be also
 some functions of $x$ as demonstrated below)
 then it follows that  the necessary  property of complete SS  is $k_0=0$.
To demonstrate it suppose the opposite: for an arbitrary $N_1,N_2$
 one can apply
 $k_0>0$.  Then $w_s$ is the sum of
two distributions $w^m$ and $w_n$ with the norm $k_0,1-k_0$.
In this case $w_s$ distribution corresponds to the probabilistic
mixture of two ensembles  with distributions $w^m,w_n$.
But this contradicts to $L_p$ proposition which exclude
any presence of  $|g^m\}$   which  characterized by $L_m$
and produce $w^m$. Now let's find how this fact restricts $w_s$
form. For the simplicity suppose that $w_{1,2}>0$ for all $x$
alike (\ref {CC}) and so $w^m>0$. Then to exclude $w^m$ admixture in $w_s$
it's enough to demand that at least in one point $x_0$ in which
$w^m(x_0) > 0$ one have  $w_s(x_0)=0$.
It seems  a  very important $w_s$ property which
shows that $m$ fuzziness can deform $w^m(x)$ form
 significantly and reveals in fact FM nonlocal, nonlinear features.
Below more strict $w_s$ properties will be found and in particular  
it will be shown that
 $w_s$  should oscillate around $w^m$ and have many
poles $x_j$.

After this simple example let's regard more subtle SS criteria - SC
 for $w_s$ which will be used throughout our formalism.
As in the first case let's decompose $w_s$ into 
the additive and nonadditive parts :
$$
  w_s(x)=k_{w1}(x) w_1(x)+k_{w2}(x) w_2(x) +w_n (x)
$$
where $k_{wi}$ are an arbitrary positive functions and  $w_n$
defined via $w_I, w_i$ analogously to above example.
 In this case we demand again $min(w_n)\ge0$  for any $x$
characterizing $w^m$ admixture of any form.
Obviously
the largest SS - sources smearing achieved for the
approximately equal signal intensities:
$ N_{1,2}=\frac{1}{2}\pm \epsilon$
at $\epsilon \rightarrow 0$
 where $F_I$ of (2) has its maximal value.
From that  SC
can be formulated as follows : in this limit in
$w_s$ decomposition  one should get
$k_{wi}(x) \rightarrow 0$, so admixture of additive $S_i$ signals
is negligible.

We argue that such  SC corresponding to
 maximal SS for $w_s$ stipulated by two
interrelated conditions.
 The first necessary condition
  is the complete $w_{1,2}$ overlap $C_x=1$
which means $w_1=w_2$. The second condition
assumes  that in FM  for any $w_i(x)$
the corresponding SS  class of functions $F_c$ exists  which permit to
achieve maximal SS.  To demonstrate the first condition meaning
 note first that
if in some  interval $Dx$ $w_2(x)=0$ and $w_1(x)>0$ (or vice versa)
then in this $Dx$ interval
 the signal from one slit in $w_s(x)$ 
 presented only i.e. $k_{w1}(x)=1$ and $w_n(x)=0$
 which excludes maximal SS achievement. 
If $w_1>w_2$ (or vice versa)
 in some $Dx$ then $w_s$ in this bin is also unambiguosly
affected but its proof is a bit more complicated.
For example for $F'_I$ one obtains:
$$
   k_{w1,2}=1-\frac{2\sqrt{w_1 w_2}}{w_1+w_2}
$$
and $k_{wi}>0$ in any $x$ where $w_1 \neq w_2$ which contradict to maximal SS.
 It's easy to check that 
 $w_1(x)=w_2(x)$
   for TSE  can be achieved at
 $l_y\rightarrow \infty$, when $w_i=const(x)$. 
 In general to find this SS  $F_c$ class
 is quite complicated problem, but for our study
 it's enough to consider   several simple cases.

 From $w_s$ Fourier transform  analysis it follows that 
  for $w_{1,2} \rightarrow const(x)$
 independently of $F_I$ this SS class  of
functions is $F_c(x)= cos(p_fx+\varphi_f)$  with an arbitrary parameters
 $p_f,\varphi_f$. Really in this $w_i$ limit for any $F_I$ it results in
 $F_I \rightarrow const(x)$ also and for such $F_c$ it follows
$\int F_I F_c dx\rightarrow 0$ which  is the normalization 
condition on $w_s$ of (2) 
 oscillations. Also
 this oscillations  have the maximal amplitude
compatible with $w_s \ge0$ and $w_s(x_i)=0$ for the infinite
number of $x_i$ poles if $F_I=1$ under condition  $w_1=w_2$.
$w_s$ dependence of $p_f,\varphi_f$
  evidences that $K$ or $\bar{g}^x$
components of $g$ are necessary  in addition to $w(x)$
for the unique final state description.
It seems a quite important observation that FM results in
 oscillating $w_s$ distribution in TSE, because in general 
it demonstrates  the possibility of fuzzy $w(x)$ nonlinear evolution.
Remind that experimentally  such oscillations observed in TSE
which is one the  direct QM nonclassicallity demonstrations.

Of course to be precise
one should study $w^0$ smearing over all space (X,Y in our case) 
at given $t$
 but $w_s(x)$ at given large  $l_y$ permits to achieve
the maximal possible SS and reproduces the main features 
necessary for the comparison of our toy-model with QM.  
Despite SC was regarded in  toy-model with one fuzzy parameter
all this SC formalism for pure and fuzzy states smearing are applicable 
for arbitrary FM theory and will be used below with account of
possible $w_i$ time dependence.

We must stress that this SS - i.e. the principal smearing of initial state
is the universal FM feature and in general case of arbitrary
state evolution  the final  $w_s(x,t)$ has analogous
dependence on $w_0$ with  additional account  of possible
 $g$ time dependence. 
 TSE is just the illustrative  example which reveals 
SS features in the most simple form. In particular $w_s$ 
ansatz (2) is applicable for a wide class of nonlinear
theories.


\section {Evolution in Fuzzy Mechanics}

Now we try to develop our approach to $m$ state evolution in
 a more formal way and try to find FM evolution properties
of free $m$ motion in 1-dimension which follows from the maximal SS conditions
regarded in the previous chapter.
 Note that below in our calculations
 a generalized complex and real functions, alike Dirac $\delta(x)$  are used,
 but  their physical meaning is well understood in standard QM
and the same approach to them applied here \cite {Ber}.

 Consider that experimentalist prepares in 1-dimension
 at $t_0$ an arbitrary $m$ pure fuzzy state $|g_0\}$ with $w_0(x)$
distribution localized inside $x_a,x_b$ interval;
other $g_0$ parameters are unimportant. After that
$m$ evolves freely to some unknown $g(t)$  with $w_s(x,t)$.
As was argued above due to SS - fuzzy smearing
 $w_s(x,t)$  have such form that at any $t$ 
 it's impossible to relate $w_s(x_i,t)$ origin to a pointlike sources
in arbitrary  small bin $Dx_j$ at $t_0$. As the result $w_s(x,t)$  has
nonlinear $w_0$ dependence which calculated here.
 We'll start with considering initial $m$
pure fuzzy state $g_0=g^0_1 \oplus g^0_2$ at $t_0$ localized in two such
small bins
i.e. $n_s=2$   : 
\begin {equation}
  w_0=w^0_1+w^0_2=N_1 \delta (x-x_1)+N_2 \delta (x-x_2)  \label {D1} 
\end {equation}
with an arbitrary $x_1,x_2$ and $N_1,N_2>0$; $N_1+N_2=1$.
This set-up is in fact 1-dimensional TSE analog and for it
$w_s$ ansatz (2)  and correspondingly
 first SS condition $w_1=w_2$  should hold also  at any $t$ independently
of $|x_1-x_2|$ value. From $x$ invariance 
it follows that for each  individual source 
 $w_i(x)=f_w(t)const(x)$; the time factor $f_w$ 
has no principal meaning  \cite {Fey}.
 Really for each realistic $m$ source $S_i$ 
   $w_i(x)$ centered around $x_i$ (or correlated with it) so  as the
result for any pair of them the overlap
$C_x<1$  and only $w_i(x)=const$ avoids it giving $C_x=1$.  
Remind that for TSE such $w_i$ achieved in the limit $l_y\rightarrow \infty$.
Yet as was shown above for the
superposition of two substates $g_{1,2}$
with such constant space distributions the corresponding  $F_c$ class 
resulting in maximal SS is :
$$
   F_c(x,t)=\cos \beta_{12}=\cos[p^f_{12}(t)x+\varphi_{12}(t)]
$$
 with an arbitrary functions $p^f_{12},\varphi_{12}$.
 $w_s$ ansatz (2) is applicable here also
and in its terms:
\begin {equation}
  w_s(x,t)=w_1(x,t)+w_2(x,t)+w_{12}(x,t)
=f_w(t)[(N_1+N_2)+F_I(N_1,N_2) F_c(x,t)] \label {DD2}
 \end {equation}
 We don't obtain
 at this stage  $F_I$, except its agreement with (2)
  but it will be calculated below. The only
its established feature is its maximum value $F_I=1$ at 
$\|w_i\|=N_i=\frac{1}{2}$ which  follows from SC conditions and
in particular it  gives  periodical $w_s(x^f_i)=0$ .
Note that this SC are correct for the pointlike sources with
the width $Dx \rightarrow 0$ but aren't applicable for the sources
 of arbitrary width.

Let's discuss  the  obtained intermediate results.
First, up to undefined at this stage $F_I,k_f,\varphi$ this ansatz for
 $w(x,t)$ coincide with the corresponding QM path integral
calculations and it doesn't
seems to be just occidental coincidence \cite {Fey}. As was mentioned above  
the meaning of $\delta(x)$  and $const(x)$ distributions formulated in
QM formalism as the spectral decomposition  of standard $L^2$
 functions \cite {Ber}
and in our theory it will be  used in the same sense. Despite, the physical
meaning of this statements will be analyzed in some limit below.
 For example
 one can regard results for $\delta(x)$ sources as the limit for the source
$S_G$ of  the initial gaussian form $w \sim e^{-\frac{x^2}{\sigma^2}}$  for
 the limit  $\sigma \rightarrow 0$. When one can claim that in this limit
$w_s$ smearing of two such sources tends to be complete in the
final state at any $t$.   
Analogously the constant $w$ distribution in $S_G$ final state
 can be approximated as
gaussian with infinitely large dispersion $\sigma \rightarrow \infty$
Hence all the following results should be regarded  as the
analogous asymptotic propositions which are exact in the suitable limit.
Eventually if  $\varphi, p^f, f_w, F_I$ will be found 
and  the consistency of such theory proved
  one  can obtain via the spectral
decomposition an evolution of any initial $g^0$ and
 the corresponding  $w^0(x)$.
To derive this functions let's regard the same set-up for three analogous
pointlike sources $n_s=3$ at $t_0$ to which
corresponds the general  $w_s$ ansatz :
\begin {equation}
   w_s(x,t)=\sum w_i(x,t)+\sum^3_{j=1}\sum_{i<j}^3 w_{ij}(x,t)+w_{1,2,3}(x,t)
 \label {DDA}
\end {equation}
which is the generalization of (\ref {DD2}) with the new triple term
$w_{1,2,3}$ which can depend on all three $w_i$.
Yet SC ansatz for $n_s=2$ should be fulfilled for each pair of sources
$i,j$ ($2,3$ for example) resulting in their maximal SS
  independently of other sources presence; otherwise
fuzzy smearing will become incomplete for $n_s\ge 3$.
$w_{1,2,3}$ presence  should deform this ansatz with the rate
 depending on $w_3(x)$.
 From that we conclude that $w_{1,2,3}=0$ and  any $w_{ij}$
ansatz coincides with $w_{12}$ of (\ref {DD2})
(here and below $i<j$ in double indexes). The same
arguments are valid for an arbitrary $n_s \ge 3$ which permit to
generalize in the obvious way  $w_s$ for $n_s=3$.  
Such $w_{ij}$ can be negative at some $x$ and thus for $n_s \ge 3$
 without the additional
constraints on $p^f_{ij},\varphi_{ij},F_{I}$ one finds that
 $w_s$ can become negative which is nonsense. 
It's argued here
 that the only solution for $w_s$  which is equivalent to this constraints is :
\begin {equation}
   w_s(x,t)=| \sum_{i=1}^{n_s} \Phi_i (x,t) |^2 \label {DD4}
\end {equation} 
where $\Phi_i$ are some $N_i$ dependent  complex functions.
We omit the detailed proof which is quite elementary but tedious
and indicate only its main points.
Let's denote as  $\beta$ constraint the equality 
 $|\sum \beta_{ij}(x)-2n\pi|=0$. As the first step we
find $w_s$ for  $N_1=N_2=N_3$ fixing $F_I$
 and consider only $p^f_{ij},\varphi_{ij}$ variations.
 It follows   that for an arbitrary 
$p^f,\varphi$ even  the infinitesimal
 $\delta p^f_{ij},\delta \varphi_{ij}$ variations results in $w_s(x)<0$ in some
$x$ interval   if  due to this variations
 $\beta$ constraint violated.
From the comparison with  $w_s$ of (\ref {DDA})
   $\beta$ constraint  for equal $N_i$ results in :
$$ 
     w_s=|\sum^{n_s}_{l=1}  \sqrt{N_l} e^{i\gamma_l(x,t)}|^2
$$
 with $\beta_{ij}=\gamma_i-\gamma_j$. From   $\beta$ constraint 
and FM  $x$ coordinate  invariance 
$\gamma_l$ can be derived up to $2n\pi$:
$$
  \gamma_l(x,t)=\eta(t)(x-x_l)^2+\alpha_l(t)  
$$
where $\eta,\alpha_l$ are an arbitrary real functions.
After that due to  $F_I,F_c$ factorization in (2) 
we can use the obtained
relations and regard $F_I$ parameters variations.
After the analogous calculations one obtains $F_{I}(N_i,N_j)=2\sqrt{N_i N_j}$
and all this results can be extended on arbitrary $n_s$ copiously.

The simplest  consistent 
$|g\}$ representation in this case  is a complex function \\
 $g(x,t)=\sum \Phi_l(x,t)$ 
  which  corresponds to $w_s(x,t)$ of (\ref{D1}).
If an initial state for $n_s=2$
 to rewrite also via $\Phi_l(x,t_0)$ one obtains :
$$
  g_0=\sum^{n_s} \sqrt{N_l}\delta(x-x_l)e^{i\alpha^0_l} 
$$
where $\alpha^0_l=\alpha_l(t)$ are  an arbitrary real constants.
Also it gives:
$$
  \eta(t)=\frac{im}{2(t-t_0)};\quad\quad
  f_w(t)={\frac{m}{2\pi(t-t_0)}}
$$
This ansatz can be extended  on arbitrary $n_s$
and from this relations $p^f,\varphi,\gamma$ easily restored; in particular:
$$
  p^f_{ij}(t)=-i\eta(t)(x_i-x_j)
$$
Thus an $n_s$ initial state  besides $N_l$  depends only on the $n_s-1$
parameters $\alpha_l^0-\alpha_i^0$ which defines the correlations
between the fuzzy sources.
If we admit that for $g_0$ one can transfer from the sum of
 $n_s\rightarrow \infty$
to an integral on $x$  smoothly then it describes the spectral
 decomposition of an arbitrary complex $g_0(x)=\sqrt{w_0(x)} e^{i\alpha(x)}$
 and its evolution for a free $m$ state : 
$$
 g(x',t)= \int G(x'-x,t-t_0)g_0(x)dx
=\int e^{\frac{im(x'-x)^2}{2(t-t_0)}} g_0(x)dx
$$
which coincides with the free $m$ evolution in QM path integral
formalism and $G$ is free Feynman propagator \cite {Fey}.
Summing up  we conclude that the obtained
 ansatz coincides with QM for the free $m$ evolution
 and in particular reveals FM state $g$ evolution linearity. 
$|g\}$ is the normalized vector (ray) of complex 
Hilbert space $\cal{H}$ which corresponds to our 
set $M_s$.


Now we need to formulate Hamilton formalism for FM. 
 For complex $g(x)$ from $x$ invariance of free $m$ motion
 the momentum can be only the Hermitian operator
 $p=i\frac{\partial}{\partial x}$ with $[\hat{p},x]=i$,
 so that  $\bar{p}=\frac{\bar{\partial\alpha}}{\partial x}$ \cite {Schiff}.
 Yet we know that  the linear complex functions evolution described by
 Schrodinger equation (SE) of QM, where Hamiltonian $H$ becomes Hermitian
operator.  It guarantees
also $m$ flow conservation and restores classical limit for
arbitrary $H$ \cite {Schiff}. From the above arguments the free
Hamiltonian follows $\hat{H_0}=\frac{\hat{p}^2}{2m}$ and
 the natural FM generalization
for the $m$ potential interactions is :
 $\hat{H}=\hat{H}_0+U(x)$.

 It's interesting to check if the maximal
SS condition conserved in the presence of interactions.
 As the example we regard the harmonic
oscillator $U=\frac{\omega_0^2 x^2}{2}$ and consider the evolution of the
initial pure state $g_0$ consist of two pointlike sources. This state
at $t>t_0$ calculated by path integral formalism. We omit here the
detailed  results which can be
reproduced easily  \cite{Fey} and
  just notice that
$w_s$ in this case coincides with the free $w_s$ of (\ref {DD2})
but with the different $f_w(t),p^f(t),\varphi(t)$. It evidences
that maximal SS principle conserves its validity
even in the strong potential field but to prove it finally
needs  more calculations for an arbitrary $U$.

The correlation tenzor of minimal FM model corresponds to
a quantum  phases differences $K(x,x')=\alpha(x)-\alpha(x')$
and quantum phase properties obtains quite natural explanation :
the real physical parameter is $K(x,x')$ - the fuzzy correlation between 
 $x,x'$ and $\alpha$ is its $x$
  representation  ambigous up to $2n\pi$.

 The same results as presented in this chapter
can be obtained from assuming some weak $M_s$ set properties
phenomenologically.
 Initially  $g$ was regarded as the abstract mathematical
object  and it don't assumed
$a \quad priory$ that $g$ corresponds to any
linear  array (vector) of functions on phase space.
We assume only that the substates summation is associative 
and the  substates has the selfsimilarity property.
 The calculations details can be found in Appendix.

From the demonstrated equivalence  FM and 
 standard QM the fuzzy smearing  can be described
in the spectral decomposition framework.
 Really, consider an arbitrary 
QM initial state $\psi_0(x)$ at $t_0$ with $w_0(x)>0$ in some $E_x$ interval.
 Let's select its  arbitrary substates $\psi'_{1,2}$
- $S_i$ sources localized in the small bins $Dx_{1,2} \in E_x$ such that 
$\psi'_i(x)=\psi_0(x)$ if $x \in Dx_i$, $\psi'_i(x)=0$ otherwise. 
From the standard QM calculations  follows 
that for QM free evolution for this $S_i$  at any $t$ in the limit
$Dx_i\rightarrow 0$ maximal SS achieved for $w(x,t)$.
Thus considered FM results for the pointlike sources acquires
the consistent asymptotic meaning via QM states spactral decomposition.
The same results can be obtained for QM substates of gaussian form
$w_G$ in the limit $\sigma \rightarrow 0$ which were discussed above
for FM states.

For the illustration of FM properties
it seems interesting to discuss  how
 $m$ momentum distribution
can be derived from Fuzzy Geometry directly
 without use of $p$ operator definition obtained
 form FM above. As was mentioned already
 it supposed that in FM relation $p=mv_x$ in one dimension
is true in a sense that
$w^v(v_x)=w'(mv_x)$ and velocity defined by relation: 
 $\bar{v}_x=\frac{\partial\bar{ x}}{\partial t}$
Consider again  the state $|g\}$ which can be presented as sum of two substates
 $|g_{1,2}\}$
which don't cross on $x$ axe : $w(x)=w_1(x)+w_2(x); \quad w_1 \cap w_2=0$, but 
their independent  momentum distributions $w'_1,w'_2$  cross  on
 $p$ axe.
 Considering relation between $w$ and $w'$   in our approach 
$w$ can be formally defined as the 'source' of $w'$
 with the relation 
for each $w_i$  $w'_i(p)=N_i w^{rp}_i(p)$ where $w^{rp}$ is the normalized
$p$ distribution.
 If for example  this is the mixture
$|g^m\}$ it gives   :
$w'_s=w'_1+w'_2$
is additive on $\|N_i\|$ analogously to CSM and 
don't result in any constraint between $w,w'$. For a pure fuzzy $|g\}$
from  $L_p$   and the  arguments given above :
\begin {eqnarray}
     w'_s(p)=w'_1(p)+w'_2(p)+w'_I(p)  \\ \nonumber
  w'_I=2 N_1^{\frac{1}{2}}*N_2^{\frac{1}{2}}* F'_c(V_g, p)
                                                           \label {CC6}
\end {eqnarray}
where $V_g$ are other $g_{1,2}$ parameters.
Let's consider  again the regarded 1-dimensional set-up with two pointlike
sources with the initial $x$ distribution (\ref {D1}).
Suppose that an experimentalist measures only $p$ (or $v_x$)
at $t_0$  without 
acquire any $x$ information. In QM it's permitted by its axioms but in FM
 one should be careful with such assumption. Then 
$w'_s(p)$ should have such nonadditive form that it should be impossible
to ascribe its components to the particular source.
From the analogous chain of arguments as for $w_s(x)$ but now
applied for the same time $t_0$  it follows that 
 for the single source $w'_i(p)=const(p)$
and 
$$
   w'_s(p)=f^p_w[N_1+N_2 +2\sqrt{(N_1 N_2} cos (d_p p+\varphi_p)]
$$
with the arbitrary $f^p_w,d_p, \varphi_p$.
Thus QM momentum  distribution derived without addressing to
$p$ operator ansatz.  
 For the complex $g(x)$
it corresponds to commutation relation:
 $[\hat{p}x-x\hat{p}]g=ig$ which
 leads to Heisenberg uncertainty relation 
$\sigma_x \sigma_p \ge 1$ for $w,w'$ which can be regarded as the weak 
(inequality) constraint. This consideration seems to us
important because it indicates that the
origin of the commutation
 relations lays in Foset properties of the phase space.

Thus QM corresponds to FM  and $g(x)$ corresponds to
$\Psi(x)$ Dirac state vector in $x$ representation.
Such theory is the simplest nonlinear dynamical theory
 with 'square root' $w^Q$ nonlinearity,
which expressed via linear complex $g(x)$ evolution.
It's important to note that SE only isn't enough 
 and one needs to define  the relation
between $|g\}$ and $w^Q$  distributions
 to construct the  complete physical theory.
Extension of this FM theory on $3$ dimensions is straightforward
and we omit its consideration here.

\section {Discussion}
In this paper we've  demonstrated that QM and mechanics on fuzzy manifold
- FM has the close correspondence at least in 1-dimensional case.
Proposed FM model contains some phenomenological statements
which is inevitable for the theory dealing with continuous spectra.
Our theory by no means aimed to disprove QM in its contemporary
form, mainly we try to
develop adequate language to describe the quantum phenomena, and for that 
purpose Fuzzy Geometry seems appropriate.
Obtained results indicate that FM
predict effects analogous to QM interference 
induced by fuzzy smearing or SS - 
sources indistinguishability  which is
the principal property of fuzzy states absent in Classical Theory.

Note that QM in this  approach seems has some formal
analogy with
classical constrained dynamics \cite {Ish}. In such theory 
 $n_e$ - number  of effective dynamical coordinates $q'$ equal to number
 $n_m$ of 
formal coordinates $q$ minus number $n_c$ of theory  dynamical constraints.
 In CSM with
constraints its formal state $P(q,p)$ expressed  by means of some linear
operators via
effective state $P'(q',p')$ with less number of degrees of freedom (DF). 
Analogously to it $w^Q$ set in FM  reduced to single complex
function $g(x)$
 due to 'fuzzy constraints'.
 Of course this is just distant
analogy and QM has more intricated structure, but it illustrates
some its properties.
In FM constraints  are weaker then in classicaL constrained dynamics,
so that no DF can be removed. Mainly they connect observables related
 to the same DF alike $x,p$ resulting in constraints inequalities
of the kind $\sigma_x \sigma_p \ge 1$.
Proposed FM formalism pretend to explain the nature of QM commutation relations
as originating from  SS property of evolution on fuzzy manifold.

In fact the need to use  complex Hilbert space for quantum states results
from  observation of $m$ interference nonlinear on initial $w^{0}(x)$.
 It can't be principally described by linear positive states alike CSM.  
 In FM approach QM state vector $|\psi\rangle$
 meaning as the mathematical but unphysical object  demonstrated.
The true physical entities are $w^Q$ distributions observed experimentally.
$|\psi\rangle$ is the mathematical tool
  which simplify  calculations of $w^Q$ evolution
from initial to final distributions which difficult to  perform directly
due to their essential nonlinearity.  
 Note that we  used $\hbar=1$ Planck  constant calibration  throughout
our formalism as it's done in  relativistic QM. But it's clear that this 
Planck constant simply connect $x,p$ scales of phase space and has no
separate meaning.

Standard QM postulates that particle state described by Dirac
state vector in Hilbert space with corresponding evolution equation.
From that it derives uncertainty of $x,p$ and $m$ paths , their interference,
 etc. On the opposite Fuzzy theory admits that $x$ coordinate and consequently
 $m$ path are principally fuzzy and from this axioms we attempt to prove
 that  $m$ states set $M_s$ is complex Hilbert space with Schrodinger
  evolution formalism. Standard Schrodinger quantization substitutes
classical pointlike particle by the new object - Dirac  state vector
conserving classical space-time structure. FM in fact takes
the opposite approach : particle stays to be material 'fuzzy' point
, but space-time set structure changed to Foset.

Note that one of unsolved problem in this FM formalism is
the nature of physical RF space coordinates - i.e. 'target space'
relative to which we describes our states $g(x)$. We associate
RF with some massive object which interactions with other objects
- solid rods, photon bunches defines geometric ordered points $x_i$.

Clearly our FM approach has the close relation to von Neuman algebra and
Quantum Logics, especially in its fuzzy sets - i.e multivalued
 logics formulation \cite {Pyk}. In this theories the set structure
of phase space is ordered one, but Logics of propositions for this space 
 is multivalued  with at least 3 outcomes : $Y/N/U$.
 In our theory standard Boolean logics used, but this multivaluedness
in fact transferred into phase space Foset and implemented
inside its geometry. Yet our approach
seems  to be closer to physics, which always use some geometry for
description of objects relations and General Relativity is good illustration
of this thesis.

FM ideas seems to be related to Quantum reference Frames theory
developed by Aharonov  and Kaufherr. They
 have shown that in nonrelativistic
Quantum Mechanics (QM) the correct definition of physical reference 
frame (RF) must differ from commonly accepted one, which in fact was
transferred copiously from Classical Physics $\cite{Aha}$. The main reason  
 is that to perform exact quantum description
 one should account the quantum properties 
not only of studied object, but also RF, despite the possible practical
 smallness \cite {Mayb9}.   
 The most simple  of this RF properties is the existence of Schroedinger
 wave packet of free macroscopic object \cite {Schiff}.
Remind that  physical RF $F^0$ is normally
associated with some macroscopic object $M$ which can perform measurements
of studied particles, for example it can be satellite in outer space.
$M$ is binded system of atoms each of them obeys to QM laws - i.e. 
evolves according to Schrodinger equation (SE). It follows then
that $F^0$ c.m. motion
also obeys to SE relative to  any other RF - $F^1$ 
and $F^0$ quantum state will be localized wave packet
with dispersion $\sigma_x$ \cite {Schiff}.
 It  introduces
additional uncertainty into the measurement of object space coordinates
 in $F^0$.
Furthermore  this effect  results in the states transformations
between two such RFs which includes quantum corrections to the
  standard Galilean group transformations $\cite{Aha}$.
In their work Aharonov and Kaufherr formulated Quantum Equivalence 
Principle (QEP) in nonrelativistic QM  - all the laws of Physics are invariant
 under transformations between both classic and this finite mass RFs
which called quantum RFs.  QRF theory if to take it seriously 
prompts to reconsider QM foundations related to space-time structure.
Really QRF existence suppose that space coordinate is the subjective entity
 connected with the object regarded as RF.

Let's regard briefly some possible consequences of QM equivalence with
regarded Mechanics on fuzzy manifold - FM. First, one can use
space-time Foset structure at quantum level, which possible
 can be applied in some specific 
 problem, alike Gravity quantization.  
Here we studied  the fuzzy phase space structure, but in relativistic
case also space-time structure must be modified analogously.
It will be interesting to compare it with Noncommutative Geometry
approach to space-time at small scale \cite {Con}.

\appendix
\section {Appendix: \quad Fuzzy states Decomposition}

Here we regard the complementary derivation of FM results
from the assumption of some simple fuzzy states properties.  
 Note that initially  $g$ was regarded as the abstract mathematical
object  which properties must be found and we don't assume
$a \quad priory$ that $g$ corresponds to any
linear  array (vector) of functions on phase space.

We define formally that
 $|g'\}$ is substate, i.e. the object  which can be operated
analogously to $|g\}$, but in general
don't describe $m$ state completely.  $|g'\}$ is linear in $M_s$ if
it can be presented
 as $|g'\}=r|g^0\}$ for $0\le r\le 1$;  for some arbitrary
state $|g^0\} \in M_s$,
 but this substates linearity is too strong condition
which means in fact $g \in M_L$ and isn't necessary.
The necessary feature is the weaker condition : $|g'_1\}$
 norm $N'_g\le 1$ and for any $Q$ for $|g'_1\},|g^0\}$  we have : 
$w^Q_1(q)=N'_g w^Q(q)$; for strong condition it corresponds
to $N'_g=f_N(r)$.
We'll call two substates $g'_{1,2}$ parallel if for them any $w^Q_{1,2}$
differ only for their norms.
 Thus $g'_i$ is parallel to a separated state $g_i$. 
 We'll regard more formal 
substates definition below, but here it's enough to use
this semiqualitative description.
An analogous substates superpositions can be defined like states
in noncrossing intervals $[q_i,q_j]$ for other $Q$ observables.
In QM a substates introduction related closely with $\Psi$
 definition as a vector
in the linear Hilbert space but in FM it results from the
 from  $m$ properties as a fuzzy point.
We assume also that substates summation is associative :
$$
  (g_1\oplus g_2)\oplus g_3=g_1\oplus (g_2\oplus g_3)
$$
For illustration we'll suppose first also substates linearity,
 but afterwards perform the same calculations dropping this assumption
conserving only the more weak property of states selfsimilarity.

Regarded TSE toy-model features supposedly  demonstrates that particles $m$
states evolution is nonlinear for $w(x,t)$ and for others $w^Q$.
From TSE analysis FM formal analogy with with  waves
evolution is straightforward : in both cases one has the flow conservation
and strong interference effects and this analogy should be explored
in detail.  
Due to well-known problems with nonlinear evolution
the wave theory results prompts us to
 look for the special $|g\}$ representation which
obeys to linear evolution equations and reproduces $w^Q$
 by some  nonlinear relations $w^Q(q,t)=F^Q(g,q,t)$.
In this linear case for  TSE   follows that
the state on the photoplate  presented as
$|g_f\}=|g_{1f}\}\oplus|g_{2f}\}$  where $g_{if}$ produced by $g_i$
and $w_s=F(g_f)$.

Now suppose that 
a fuzzy states  admits selfsimilar decomposition. It
 means that any fuzzy state $|g\}$ can be decomposed identically
into the system of parallel to $g$ substates $|g_i\}$. Moreover any such
 substate in its turn can be decomposed into another substates system.
For the start here we'll assume  parallel substates linearity - i.e.
strong condition formulated above, but after that
 calculate it for weak condition which don't suppose $g$ linearity.
Consider  arbitrary fuzzy  state $|g\}$, obviously it can be rewritten as :
$$
   |g\}=|g_1\}\oplus |g_2\}=s_1|g\}\oplus s_2|g\}
$$
with $0\le s_{1,2} $  and $s_1+s_2=1$.
 This equality 
supposedly fulfilled  for our undefined $|g\}$  summation rule, whatever it is.
Note that such equality holds for quantum state vectors, yet calculating
corresponding QM probabilities one should formally account appearing IT,
which is the essence  of the regarded ansatz.
 For parallel substates   $w_i(x)=f(s_i)w(x)$ for any
monotonous $f$, so we can also use $f$ value to identify substates. 
Interference of parallel substates obviously can be only constructive
 and independent of $x$, so $F_c(x)=c_g \ge 0$. 
 For the lack of place we don't
prove it here, but following calculations illustrate it.
Let's rewrite  (2) for this substates and obtain from it by
 integration over $x$ the
relation for their norms. If to denote $u_0=1; u_1=s_1; u_2=s_2$
 and remind that $f(1)=1$  after canceling $\|w_s\|$ on left 
 and right side one obtains :   
\begin {equation}
  f(u_0)=f(u_1)+f(u_2)+F_I(f(u_1),f(u_2)) F_c \label {CC4}
\end {equation}
This equality is supposedly true also for substates
 and if we decompose $|g_2\}$  analoguosly :
$$
   |g_2\}=|g_3\} \oplus |g_4\}=s_3|g_2\}\oplus s_4|g_2\}
$$
with $s_{3,4}\ge 0 ; s_3+s_4=s_2$
 then equality (\ref {CC4}) must be true for substitution
$u_0=s_2; u_1=s_3; u_2=s_4$. This equalities must be hold simultaneously
for arbitrary $s_i$ inside described intervals and constitute equations
system  which permit to derive
$f,F_I$. Assuming that no new dimensional parameters appears,
after simple algebra omitted here  obtain: \\
$$
   f(z)=z^n; \quad c_g=1 ;\quad 
   F_I(z_1,z_2)=(z_1+z_2)^n-z_1^n-z_2^n;
 \quad n=1,2,...\infty 
$$
Now we drop a substates linearity assumption
 changing it to weak condition for norm
and use as substate $|g_2\}$  parameter $f=f(s_2)$,
so admit $w_2=fw;\quad w_1=f'(f)w$ and $f+f'(f)\le 1$.
From it  and (\ref {CC4}) we obtain equality: 
$$
  1=f+f'(f)+F_I(f,f') F_c
$$
Then  again decomposing $|g_2\}$ into $|g_{3,4}\}$ from corresponding
equalities system the   solution obtained :
\begin {eqnarray}
   f'=(1-f^{\frac{1}{n}} )^n; \quad c_g=1 ; \quad \\ \nonumber
   F_I=(f^{\frac{1}{n}}+(f')^{\frac{1}{n}})^n-f-f'; \label {CC99}
 \quad n=1,2,...\infty 
\end {eqnarray}
So even without substates linearity we get essentially the same results
which restricts $F_I$ choice severely.

Here $n=1$ corresponds to classical probability without IT 
i.e. to CSM. $n=2$ possibly corresponds to QM
 and classical waves interference
and so deserves special attention. Note that $F_c=c_g$ only for parallel
substates, but it can be also negative for destructive interference and
so for arbitrary substates $-1\le F_c \le 1$.
 Yet for $n \ge 3$ for $F_c \le 0 $ 
$w_s$ becomes negative which principally contradicts to its definition.
In general this solutions exists also if  $n$ changed to rational
 $\frac{j}{l}$ or continuous $y_n$, but it doesn't bring something new in
our arguments and we'll regard  only natural $n$.

Yet by use of associativity even for weak condition without $g$ linearity
by renormalization of given states one can obtain
new  states $g^r$ which obeys to linearity.
For example for $n=2$ one can substitute $s'_1=\sqrt{f};\; s'_2=\sqrt{f'}$ 
and it follows $s'_1+s'_2=1$. Thus we can express $|g^r_i\}=s'_i|g\}$
and in this case strong and weak conditions on substates coincide,
below we regard $g$ and $g^r$ states on the same ground. 

 Until now we operated only with real
and  not complex values  which will be regarded now. 
Consider for  $n=2$ and $|g_{\alpha}\}=e^{i\alpha}|g\}$ for real $\alpha$
 if to settle $f=|s|^2$ for complex $s$ then
$|g_{\alpha}\}$ describes the same distribution $w(x)$.
Really any $w^Q$ depends on $g$ internal structure and can't
change after such multiplication.
Let's return to $|g\}$ selfsimilar decomposition, but permit
$s_1,s_2$ to be complex : $s_i=d_i e^{i\alpha_i};\;d_i\ge 0$.
Conserving $s_1+s_2=1$ constraint 
now for $u_0=1; u_i=s_i$ eq. (\ref{CC4}) fulfilled 
with $f,F_I$ defined above. 
Now we can't claim $F_c=1$ because interference of substates with
complex $s_i$ can't be guaranteed to be constructive.  
From simple calculations if to settle that
the complex component of (\ref {CC4}) equal to $0$ 
one obtains that $F_c=\cos(\alpha_1-\alpha_2)$

Obtained results  deserves some comments.
In standard QM such calculus don't play any role, because IT form
defined by  Hilbert space properties introduced $ad \; hoc$, in particular
the scalar product definition.
In our model we defined only minimal  $M_s$ properties, and thus
IT form is arbitrary and is derived or at least constrained
 assuming $M_s$ fuzzy states selfsimilar structure.
The selfsimilarity means that the fuzzy states
introduced consistently and in particular any substate $g_i$ has all the
formal properties
of complete $m$ state. We can't exclude completely the theories 
with $n \ge 3$ now, but at least
they seems to include quite intricated nonlinear effects.
 Such selfsimilarity
is quite often meet in  other theories and so can have universal meaning.
Note that the nonunique  norm and distance  definition permitted
for complex functions spaces $M_L^p$ for $P\ge 1$ \cite {Edw}, but
only for $P=2$ scalar product can be defined and $M_L^p$ becomes
Hilbert space. Our calculations leads to the analogous result : 
$n=2$ is preferable for consistent theory
, but from different arguments and this correspondence
deserves further study. Anyway, our ansatz at $n=2$ has
 $|g\}$ scalar product definition.

To get this results we admitted only $|g\}$ state decomposition into
parallel substates which observed properties identical to $|g\}$
except the norm.  Earlier
we assumed that $|g\}=\sum\oplus|g_i\}$ sum of substates in noncrossing
regions $Dx_i$ on which effective $X$ surface can be decomposed.
In the limit of $Dx_i\rightarrow 0$ it can be expressed
as $|g\}=\int \oplus |g_x\}dx$.
Now we   regard more restrictive  hypothesis
 that $|g\}$ has $x$-representation i.e.
fuzzy state described completely via $\vec{g}_x=\{g^{\mu}(x)\}; \mu=1,n;$    
countable set (vector) with arbitrary $n$  number of
  real or complex functions components, plus scalar product definition.
 Consequently summation rule $\oplus$
describes component sum and states set is  $M_L$ linear functions space. 

This $x$ representation corresponds well to fuzzy geometry
which supposes that  $m$ evolution features can be encoded
in some functions on $X$ axe.
 Turning back to $g_F=\{w,K(x,x'\}$ structure regarded in chap.3  
note that it admits $x$-representation if $K$ can be presented
as $K=K_F(x)-K_F(x')$  which will be obtained here eventually.

Now we can apply the selfsimilar decomposition
in any point $x$ and $s_i$ can be arbitrary functions. 
 Eq. (\ref{CC4}) now is applicable in any point $x$
for $w_s(x)$ and $s_i(x)$  with the same
constraint $s_1+s_2=1$. It follows for complex
 $s_i(x)=d_i(x) e^{i\alpha_i(x)}$ for arbitrary real functions $d_i,\alpha_i$
that it should be
 $$
    F_c(x)=\cos(\alpha_1(x)-\alpha_2(x))
$$
 Thus if it's possible to factorize from $\vec{g}_{1,2}(x)$ states the
 complex multipliers
$g^c_{1,2}(x)$ than for two states sum $g_1+g_2$,
 $F_c$ depends on them along with  this ansatz for $s_i(x)$ and
 $ F_I=2 \sqrt {w_1 w_2}$ for $w_s$.

After this formalism development
 the general evolution problem (GEP) can be regarded:
experimentalist prepares arbitrary fuzzy state $g_0$ and 
should  calculate the  future state $g(t)$.
 $g_0$  can be presented as
the sum of  sources $w^0_i$ in small $\Delta x_i$ : $|g_0\}=\sum |g_i\}$.
Alike for TSE where it follows from $L_p$ property in general pure case 
 $m$ source doesn't attributed to any $\Delta x_i$ :
$m\notin \Delta x_i$. Due to it $g$  evolution should
smear the signals from different $\Delta x_i$ for final $w_s(x,t)$
 which should obey to SC condition and so restricts possible $g$ evolution.
From the arguments discussed for TSE :
$$
   w_s(x,t) =\sum w_j(x,t)+w_I (w^0_j,x,t)
$$ 
From selfsimilarity arguments it follows for IT :
$$
     w_I=\sum\sum  \sqrt {w_i^0 w_j^0} G(x,t,x_i,x_j,g^p_i,g^p_j)
$$
where $G$ is double correlated $m$ propagator and
 $g^p_i$ are $g_i$ parameters
different from $w^0_i$. Note that in distinction from TSE
the time dependence of $w_s$ becomes important. Of course
 other $w^Q$ distributions has the same principal properties, but for
$x$ they are most obvious. 

It's easy to show that single real function $g_r(x)$ can't
 satisfy to formulated demands for $g_x$. If we regard
the  complex function $g(x)=d(x)e^{i\alpha(x)}$ 
 with $d\ge 0$ and  $\alpha$- real then it follows $w(x)=g^* (x)g(x)$.
For TSE $w_I=2d_1 d_2 \cos \alpha_{12}$; 
where $\alpha_{12}(x)=\alpha_1(x)-\alpha_2(x)$. Thus if for
TSE model  dynamics results in monotonous $\alpha_{12}(x)$ and  fuzzy
 separation
criteria  SC fulfilled for oscillating $w_s$ then complex $g(x)$ can be good
candidate for fuzzy state. Yet we know that such dynamics supplied by
Schrodinger equation (SE) of QM, where Hamiltonian $H$ becomes Hermitian
operator. It guarantees
also $m$ flow conservation and restores classical limit for
arbitrary $H$ \cite {Schiff}.
The same is true for GEP $g(t)$ and it can be  shown
that fuzzy  SC fulfilled for arbitrary initial $g(x)$ due to
quantum interference. $w_s(x,t)$ has the form analogous to (2) but
with large number of IT terms.
It's important to note that SE only isn't enough 
 and one needs to define a relation
between $|g\}$ and $w^Q$ - experimental distributions
 to construct the  complete physical theory.
 Note than in QM
TSE oscillations only approximately described by $\cos {kx+\alpha}$,
but it doesn't disprove our arguments \cite {Schiff}.
Thus QM supposedly corresponds to FM with $n=2$ and $g(x)$ corresponds to
$\Psi(x)$ Dirac state vector in $x$ representation.



\begin {thebibliography}{99}


\bibitem{Aha} Y.Aharonov, T.Kaufherr Phys. Rev. D30 ,368 (1984)
 
\bibitem{Dop} S.Doplicher et. al. Comm. Math. Phys. 172,187 (1995) 

\bibitem{Schiff} L.I.Schiff, 'Quantum Mechanics' (New-York, Macgraw-Hill,1955) 

\bibitem {Con} A.Connes 'Noncommutative Geometry' (Academic Press, 1994) 

 
 H.Kitada Nuov. Cim. 109B , 281 (1995) , gr-qc/9708055

\bibitem {Ish} C.Isham in :'Canonical Gravity : from Classical to
Quantum ' Eds. J.Ehlers, H.Friedrich , Lecture Notes in Phys. 433
(1994, Springer, Berlin)
\bibitem {B} S.Baez et. al. Com. Math. Phys. 208, 787 (2000) 

\bibitem {Zad} L.Zadeh Inform. and Control 8, 338 (1965),
 IEEE Trans., SMC-3, 28 (1973)

\bibitem {Got} H.Bandemer, S.Gottwald 'Einfurlung in Fuzzy-Methoden'
(Academie Verlag, Berlin, 1993); English edition published.

\bibitem {Ali} T.Ali , Journ. Math. Phys. 15, 176 (1974);
ibid., 18, 219, (1977)

\bibitem {Pyk} J.Pykaz Found. Phys. 30, 1503 (2000)

\bibitem {Ren} M.Requardt , S.Roy gr-qc/0011076 , Class. Quant. Grav (2001)

\bibitem {Mayb8} S.Mayburov, Proc. V Quantum communications and measurements
Conf.,Capri, 2000 (Cluwer,2001), Quant-ph 0103161


\bibitem {Busch} P.Busch, P.Lahti, P.Mittelstaedt,
'Quantum Theory of Measurements' (Springer-Verlag, Berlin, 1996)

\bibitem {Ber} F.Berezin, 'Shroedinger Equation' (Moscow, Nauka, 1985) 

\bibitem {Kol} A.Kolmogorov, 'Information Theory' (Moscow, Nauka, 1957)  

\bibitem {Edw} R.Edwards 'Functional Analysis and Applications'
(N-Y, McGrow-Hill, 1965)

\bibitem {Fey} R.Feynman,A.Hibbs 'Quantum Mechanics and Path Integrals'
(N-y, Mcgrow-Hill,1965)

\bibitem {Mayb9} S.Mayburov Proc. VII Marcel Grossman Conf.,
Jerusalem ,1997 (W.S.,Singapour,1998), hep-th 0007003.



 

\end {thebibliography}
\end {document}